\documentclass[%
 aip,
 amsmath,amssymb,
 reprint,%
]{revtex4-1}

\usepackage{graphicx}
\usepackage{dcolumn}
\usepackage{bm}

\usepackage[utf8]{inputenc}
\usepackage[T1]{fontenc}
\usepackage{etoolbox}
\usepackage{color}
\usepackage{hyperref}
\usepackage{tabularx}
\usepackage{makecell}
\usepackage[figuresleft]{rotating}
\usepackage{xcolor}
\usepackage{soul}

\include{vpack}

\makeatletter
\def\@email#1#2{%
 \endgroup
 \patchcmd{\titleblock@produce}
  {\frontmatter@RRAPformat}
  {\frontmatter@RRAPformat{\produce@RRAP{*#1\href{mailto:#2}{#2}}}\frontmatter@RRAPformat}
  {}{}
}%
\makeatother
\begin{document}

\preprint{AIP/123-QED}

\title{Differentiable Turbulence: Closure as a PDE-constrained optimization}

\author{Varun Shankar}
\affiliation{Carnegie Mellon University}

\author{Dibyajyoti Chakraborty}
\affiliation{Pennsylvania State University}

\author{Venkatasubramanian Viswanathan}
\affiliation{University of Michigan}

\author{Romit Maulik}
\affiliation{Pennsylvania State University}
\affiliation{Argonne National Laboratory}
\email{rmaulik@psu.edu}

\date{\today}

\begin{abstract}
Deep learning is increasingly becoming a promising pathway to improving the accuracy of sub-grid scale (SGS) turbulence closure models for large eddy simulations (LES). We leverage the concept of differentiable turbulence, whereby an end-to-end differentiable solver is used in combination with physics-inspired choices of deep learning architectures to learn highly effective and versatile SGS models for two-dimensional turbulent flow. We perform an in-depth analysis of the inductive biases in the chosen architectures, finding that the inclusion of small-scale non-local features is most critical to effective SGS modeling, while large-scale features can improve pointwise accuracy of the \textit{a-posteriori} solution field. The velocity gradient tensor on the LES grid can be mapped directly to the SGS stress via decomposition of the inputs and outputs into isotropic, deviatoric, and anti-symmetric components. We see that the model can generalize to a variety of flow configurations, including higher and lower Reynolds numbers and different forcing conditions. We show that the differentiable physics paradigm is more successful than offline, \textit{a-priori} learning, and that hybrid solver-in-the-loop approaches to deep learning offer an ideal balance between computational efficiency, accuracy, and generalization. Our experiments provide physics-based recommendations for deep-learning based SGS modeling for generalizable closure modeling of turbulence.
\end{abstract}


\maketitle

\section{Introduction}

Simulations of turbulent flow constitute an integral part of modeling and analysis for many scientific and engineering problems driven by fluid flow. Many real-world flows, such as climate dynamics, jets, blood flow, or external aerodynamics, are turbulent in nature, characterized by chaotic and multi-scale behavior \cite{Pope2000,Tennekes2018-mf}. Numerical solutions to turbulent flow are often extraordinarily challenging to obtain, due to the vast number of temporal and spatial scales that must be resolved. In practice, direct numerical simulation (DNS) of the governing Navier-Stokes equations is infeasible and hence DNS is limited to canonical flow configurations \cite{Moin1998}. Major efforts in computational fluid dynamics (CFD) methods have therefore focused on the development of turbulence models, which approximate the effects of turbulence and ultimately reduce the computational burden of simulations by averaging unresolved flow features \cite{Argyropoulos2015}. For example, Reynolds-Averaged Navier-Stokes (RANS) methods develop steady-state solutions to the governing equations by modeling the time-averaged turbulent fluctuations in the flow encompassed by the Reynolds stresses \cite{Tennekes2018-mf, Gatski2001}. The smooth solutions provided by the RANS equations enable a significant reduction in grid-resolution requirements, and RANS has remained the workhorse of CFD methods for many engineering tasks. The increase in computational capabilities over the last few decades has placed new emphasis on Large Eddy Simulation (LES) methods \cite{Sagaut2006-ln, Pope2004, Schalkwijk2015, Argyropoulos2015}. With LES, the large scales of the flow are directly resolved on the computational grid, and the effects of the subgrid-scale (SGS) flow are accounted for using an SGS model. The advantage of LES is that the temporal evolution of the flow field can be modeled, which is needed for certain applications such as weather forecasting \cite{Shen2022}. On the other hand, the discretization requirements for LES are generally more strict than RANS, leading to larger computational costs that limit applicability. 

The averaged or `filtered' equations used in turbulence modeling are near identical to the governing Navier-Stokes equations, with the exception of an effective source term that embodies the influence of the unresolved flow. The accuracy of the turbulence model therefore has a significant impact on the quality of the resulting solution field \cite{Spalart2015, Geurts2002}. Traditional turbulence models have been developed from theory and empirical testing. While these methods have provided a strong foundation for further research and improvement, novel approaches will be required to address the inherent limitations of current models \cite{Reynolds, Duraisamy2019}. Given the partially data-driven nature of modeling approaches, it is a natural extension to consider machine learning (ML) paradigms for development such as deep learning. Data-driven turbulence modeling through deep learning has seen explosive interest in the last few years, particularly for RANS models \cite{Beck2021, Kutz2017}. Ling, et al.'s seminal work outlines an approach to model the Reynolds stress anisotropy tensor with an artificial neural network (ANN)\cite{Ling2016}. Other avenues have been explored as well, including iterative methods \cite{Liu2021}, field inversion \cite{Singh2017}, and wall-modeling \cite{Bin2022}.
Simultaneously, recent efforts have been devoted to learning more accurate SGS models for LES. Maulik et al. explored learning the SGS stress using an artificial neural network (ANN) with promising results \cite{Maulik2018}. Wang et al. examined the necessary input features for an SGS model using random forests and ANNs\cite{Wang2018}. Frezat et al. modeled the SGS scalar flux and constrained their model with invariances and symmetries \cite{Frezat2021}. Furthermore, Guan et al. used convolutional networks to learn SGS closures and investigated the effectiveness of transfer learning for generalization \cite{Guan2022,Guan2023,Subel2023}.

Before proceeding further, we elaborate on the notion of the `filter' in LES. In several data-driven closure modeling studies for LES, and in particular for most a-priori \cite{Maulik2018,Guan2022} variants, optimal subgrid stress quantities are computed for the purpose of ascertaining ground truth. These are obtained through the computation of filtered flow-fields assuming a low-pass spatial filter such as a Gaussian filter or a sharp spectral cut-off filter. It has also been observed that the optimal subgrid stress is a function of this filter \cite{piomelli1991subgrid}. Several studies have demonstrated that such assumptions are unreasonably strong and result in ad-hoc clipping requirements or strong inductive biases for a-priori trained data-driven closures during a-posteriori deployments \cite{Maulik2018,fan2024differentiable,zhao2024mitigating,sanderse2024scientific}. We note that while \textit{a-priori} insights are critical to deeper understanding of SGS effects, the desired goal is to accurately reproduce the \textit{a-posteriori} flow field. Notably, \textit{a-posteriori} evaluation introduces several additional consequences that cannot be modeled with \textit{a-priori} strategies alone, including numerical and discretization errors and temporal effects. 

One approach to overcome with \textit{a-posteriori} data-driven turbulence modeling is the development of differentiable CFD solvers, which are required to backpropagate an \textit{a-posteriori} error to the parameters of the turbulence model. We term the application of differentiable programming techniques to enhance turbulence models ``differentiable turbulence''. While alternative machine learning techniques such as reinforcement learning have been explored to circumvent the need for differentiable simulations \cite{Novati2021, Zhao2020, ObiolsSales2020}, the majority of literature on ML for spatiotemporal turbulence forecasting has leveraged pure ML architectures, where the flow solution is obtained not through traditional finite-volume or finite-element numerical methods, but rather a deep learning model alone \cite{Pandey2020, Vinuesa2022, stachenfeld2022learned}. For example, LSTMs \cite{mohan2018deep}, GANs \cite{Bode2018}, and other physics-informed approaches \cite{Wang2020} have been used to model turbulent flow. The limitation of these approaches is that without explicit knowledge of the underlying physics, generalization to unseen flow configurations can be especially challenging. Despite the challenges, the adoption of differentiable programming paradigms has led to several publications of differentiable CFD solvers in the last few years \cite{Kochkov2021,Holl2020Learning,BEZGIN2022108527}. These differentiable solvers are still very much in the nascent stages of development, given the difficulties associated with writing a CFD solver from the ground up, and for now, cannot compete with the vast ecosystem of robust non-differentiable numerical solvers in terms of applicability. 
However, they offer a pathway to exploring differentiable turbulence techniques for designing hybrid physics and ML solvers that combine the learning power of deep learning methods with the generalization ability of well-tested numerical methods. However, there are also some recent examples of \textit{a-posteriori} learning methods that utilize emulators of non-differentiable solvers \cite{frezat2023gradient}.

The concept of ``reverse-mode differentiation" in CFD simulations has existed for decades and is analogous to solving the adjoint problem \cite{Jameson1995,Mller2005,Kenway2019}. Adjoint solutions have historically been used for applications such as shape and control optimization \cite{Tonomura2010,Liu2015,Rumpfkeil2008,Giles2000}, but have garnered renewed interest with the growth of machine learning. Differentiable simulations have been integrated with ML as ``solver-in-the-loop" approaches particularly for correcting and improving the accuracy of coarse-grained or unresolved simulations, which offers a balance between computational cost and physics-embedding. Previous works have targeted learning correction operators to the solution field \cite{um2021solverintheloop}, or learning interpolation functions within the numerical scheme \cite{Kochkov2021}, which have shown promising results relative to both baseline traditional numerical solution methods and pure ML models. Given that this work is specifically focused on LES, we choose to learn an existing term that appears within the governing PDE, the SGS stress, which is an interpretable and well-studied quantity that is amenable to analysis with conventional techniques. Some works have leveraged differentiable simulations for learning turbulence models, notably List et al. \cite{List2022}, who extend the \texttt{PhiFlow} solver for SGS modeling, Sirignano et al. \cite{Sirignano2020}, who use the stochastic adjoint method for LES modeling, and Frezat et al. \cite{Frezat2022}, who examine backscatter in quasi-geostrophic turbulence. Our aim here is to demonstrate the effectiveness of ``solver-in-the-loop" approaches to learning turbulence models, to offer an in depth examination of the inductive biases contained within the models, and to provide insight for future development of data-driven SGS models.

In this work, we learn SGS closure models using a deep learning-embedded differentiable CFD solver. We consider two-dimensional (2D) homogeneous isotropic turbulence as a candidate test problem to evaluate our approach and demonstrate the capability of the learned models to produce accurate \textit{a-posteriori} solution trajectories. Our novel contributions are as follows: (1) we design and test several methods to model the SGS stress tensor, inspired by existing eddy viscosity models and tensor theory, within the differentiable physics paradigm, (2) we perform a deep analysis of network architectures to understand the importance of non-local SGS models and which length scales are necessary to capture for accurate modeling, and (3) we compare our approach with offline learning of the SGS stress to validate the need for end-to-end optimization enabled by the differentiable solver.

\section{Methods}
\subsection{Governing Equations}
Turbulence models are developed for the two-dimensional incompressible Navier-Stokes equations in a doubly periodic domain, given in its non-dimensional form as:
\begin{align}
\frac{\partial \mathbf{u}}{\partial t}+\nabla\cdot(\mathbf{u}\otimes\mathbf{u})&=\frac{1}{Re}\nabla^2\mathbf{u}-\frac{1}{\rho}\nabla p+\mathbf{f} \\
\nabla\cdot\mathbf{u}&=0,
\label{eq:ns}
\end{align}
where $\mathbf{u}$ is the velocity vector, defined $\{ u,v\}$, $p$ is the pressure, $\rho$ is the density, $Re$ is the Reynolds number, and $\mathbf{f}$ represents any external forces. The dimensionless parameter $Re$ represents the ratio of convective to diffusive forces in the flow and characterizes many critical aspects of the flow behavior. At low Reynolds numbers, viscous forces dampen instabilities in the flow leading to smooth solutions with low frequency components. At high Reynolds numbers, inertial forces can amplify minor perturbations in the flow to produce energy-containing fluctuations or eddies in the solution field that span many orders of magnitude in space and time \cite{Pope2000}. If the high frequency components of the solution field exceed the resolution of the numerical grid, a direct numerical simulation of the Navier-Stokes equations can become under-resolved, resulting in discretization errors and inaccurate solutions. High Reynolds number flows can therefore be cost-prohibitive or infeasible to evaluate with DNS, given the grid spacing required to fully resolve the flow field.

In many scientific and engineering applications, it is desirable to simulate high Reynolds number flows at computationally tractable resolutions much lower that what is imposed by DNS. In such a scenario, the governing equations must be augmented to account for the unresolved length-scales in the solution field. LES resolves the flow field up to some cut-off length scale $\Delta$. Conceptually, it is assumed that the high frequency contributions are removed via a filtering operation, i.e.,  a filter $G_\Delta$ with characteristic length scale $\Delta$ is convolved with the velocity field to arrive at the target solution field:
\begin{equation}
    \overline{\mathbf{u}}(x, t) = G_\Delta \star \mathbf{u}(x,t).
\end{equation}
Consequently, to solve for the filtered velocity field, the governing equations can be represented in terms of filtered variables with the introduction of a source term in the momentum equation to account for unresolved interactions between resolved and unresolved scales in the flow-field. Therefore, we have,
\begin{equation}
    \frac{\partial \overline{\mathbf{u}}}{\partial t}+\nabla\cdot(\overline{\mathbf{u}}\otimes\overline{\mathbf{u}})=\frac{1}{Re}\nabla^2\overline{\mathbf{u}}-\frac{1}{\rho}\nabla \overline{p}+\overline{\mathbf{f}} + \nabla\cdot \boldsymbol{\tau},
    \label{eqn:les}
\end{equation}
where $\boldsymbol{\tau}$ represents the effects of the unresolved velocity components on the resolved field and is defined as
\begin{equation}
    \tau_{ij} = \overline{u_iu_j}-\overline{u}_i\overline{u}_j.
    \label{eqn:t_dns}
\end{equation}
This quantity is termed the subgrid-scale (SGS) stress and has a nontrivial impact on the filtered velocity field. Ignoring the term can lead to numerical errors or spurious solution fields, however, it cannot be directly computed without access to the unresolved velocity components. Instead, it is the role of the SGS turbulence model to approximate $\boldsymbol{\tau}$ from the resolved field such that the problem can be closed. We emphasize, again, that the notion of `filtering' a DNS field is a conceptual tool to describe the coarse-grained evolution of an LES. The nature of this filter is generally unknown, but can be crudely approximated through \emph{a-priori} knowledge of the numerical scheme (finite volume, spectral, etc.). However, for almost all practical problems, the nature of this filtering operation is unknown which complicates data-driven modeling for $\tau_{ij}$.

Solutions to the Navier-Stokes and filtered LES equations are computed using finite-volume code written in domain-specific language designed for differentiable programming, \texttt{JAX-CFD} \cite{Kochkov2021}. \texttt{JAX-CFD} takes advantage of native \texttt{JAX} autodifferentiation \cite{jax2018github} to allow users to compute gradients of any parameter in the solver using reverse-mode differentiation. The implementation uses the discrete adjoint method \cite{Mader2008} for efficiently propagating gradients through linear solves in the algorithm. The numerical scheme uses a staggered grid for the velocity and pressure fields, second-order central difference schemes for fluxes, and explicit Euler time integration.

True solutions are obtained from a high-resolution simulation of the governing equations. Target fields are computed by filtering and coarse-graining using $G_\Delta$, a Gaussian filter with width twice the grid spacing and a spectral cutoff at the Nyquist frequency of the LES grid. This choice of filter is effective for data-driven SGS modeling \cite{Zhou2019} and additionally selected based on the use of second-order numerical schemes. More details on the specifics of the datasets are provided in Sec. \ref{sec:data}.

\subsection{Subgrid-scale (SGS) modeling}

The goal of the SGS model is to estimate the SGS stress $\boldsymbol{\tau}$, whose divergence functionally appears as an additional source term on the right-hand side of the momentum equation. Over the last few decades, several approaches have been proposed to model the SGS stress. Given that the primary function of the SGS model is to remove energy from the flow to account for the transfer of energy from the resolved scales to the unresolved scales, the most widely used SGS models are dissipative linear eddy viscosity models, commonly used in RANS turbulence models as well. The eddy viscosity model postulates that the SGS stress is proportional to the rate-of-strain through an effective eddy viscosity:
\begin{equation}
    \tau_{ij} = -2\nu_t\overline{S}_{ij},
\end{equation}
where $\nu_t$ is the eddy viscosity and $\overline{S}_{ij}=\frac{1}{2}(\partial_j \overline{u}_i+\partial_i \overline{u}_j)$ is the rate-of-strain tensor. The most well-known SGS eddy viscosity model is the Smagorinsky model \cite{SMAGORINSKY1963}, where the eddy viscosity is determined through the characteristic length scale $\Delta$ and a characteristic velocity $\Delta|{\overline{\mathbf{S}}}|$, where $|\overline{\mathbf{S}}| = (2\overline{S}_{ij}\overline{S}_{ji})^{1/2}$ to give
\begin{equation}
	\begin{aligned}
		\nu_t &= (C_s \Delta)^2|\overline{\mathbf{S}}| \\
		\boldsymbol{\tau}_{smag} &= -2\nu_t\overline{\mathbf{S}},
		\label{eqn:smag}
	\end{aligned}
\end{equation}

where $C_s$ is a dimensionless empirical coefficient. The Smagorinsky model has been effectively used in a variety of application areas, however, its simplistic nature is directly tied to assumptions that can lead to deficiencies for certain flows, namely the use of an isotropic, positive eddy viscosity. The restriction of $\nu_t$ to positive values means the model is purely dissipative and cannot model the transfer of energy from subgrid to resolved scales, also known as backscatter.

The approach taken in this study is to learn an additive correction to the Smagorinsky model using a data-driven modeling paradigm to account for SGS effects that cannot be captured by the Smagorinsky model alone. The SGS stress is therefore given by
\begin{equation}
    \hat{\boldsymbol{\tau}} = \boldsymbol{\tau}_{smag} + \boldsymbol{\tau}_{ml},
    \label{eqn:sgs}
\end{equation}
where $\boldsymbol{\tau}_{ml}$ represents a tensor-valued correction to SGS stress that is computed from a deep neural network. Such an approach has strong foundations in existing SGS modeling strategies and is considered a mixed model \cite{Zang1993}, where an eddy viscosity term is added to the stress from an alternative modeling approach such as dynamic \cite{Lilly1992} or deconvolution \cite{Adams2002} methods.

\begin{figure*}
  \begin{center}
      \includegraphics[width=0.9\linewidth]{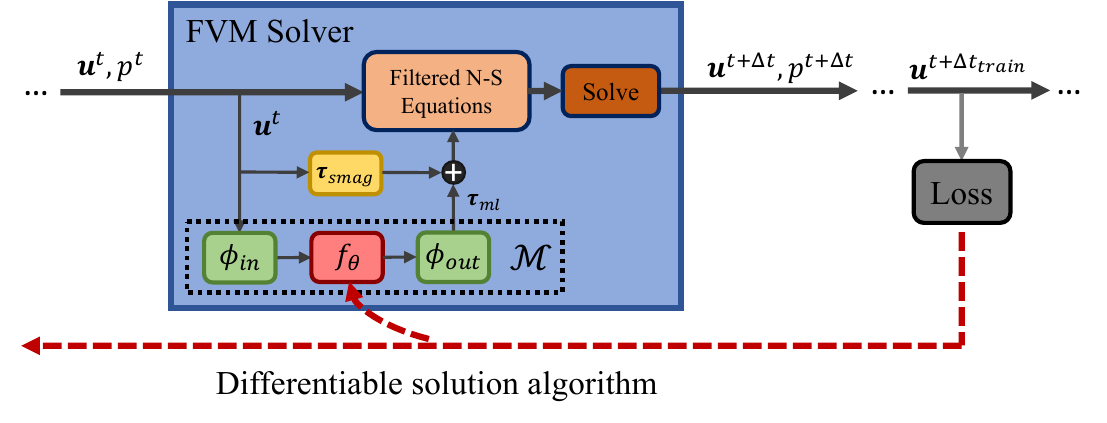}
  \end{center}
  \caption{A schematic of the deep learning-embedded solution algorithm. At each time step, the eddy-viscosity contribution of the subgrid stress is computed from $\boldsymbol{\tau}_{smag}$ and the ML contribution from $\mathcal{M}$, where $f_\theta$ represents a neural network with trainable parameters and $\phi_{in},\phi_{out}$ are fixed transformations of the inputs and outputs respectively. The contributions are summed and used in the LES equations, which are solved using standard numerical schemes. The solution trajectory is propagated and the loss is evaluated with respect to the partial (i.e., subsampled) observations of the ground truth DNS field. Because the solution algorithm is differentiable, the loss can be backpropagated through all time steps and linear solves to update the trainable parameters.}
  \label{fig:schem}
\end{figure*}

Given this formulation, the question of how to compute $\boldsymbol{\tau}_{ml}$ using a neural network remains nontrivial. We first assume that $\boldsymbol{\tau}_{ml}$ is a function of the instantaneous filtered velocity field $\overline{\mathbf{u}}$. Then, in general, the SGS model can be represented as,
\begin{equation}
    \boldsymbol{\tau}_{ml} = \mathcal{M}(\overline{\mathbf{u}}, f_{\theta}, \phi_{in}, \phi_{out}),
    \label{eqn:sgs_ml}
\end{equation}
where $f_{\theta}$ is a neural network with parameters $\theta$ and $\phi_{in}$ and $\phi_{out}$ represent transformations of the input velocity field and neural network outputs respectively. Each of the three functions can be varied to optimize model accuracy and learning. In this section, we discuss $\phi_{in}$ and $\phi_{out}$. Variation of $f_{\theta}$ will be considered in the following section. We show a general schematic of the algorithm in Fig. \ref{fig:schem}.

A naive implementation of the model would be to take $\phi_{in}$ and $\phi_{out}$ as identity maps, but this approach has several drawbacks. $\phi_{in}$ as the identity function has been commonly used in previous studies of data-driven turbulence modeling, particularly for LES \cite{List2022}, however, there is an immediately observable flaw in that the model does not respect Galilean invariance. A change in inertial reference frame would result in different network outputs and subsequently $\boldsymbol{\tau}_{ml}$. A straightforward solution is to start by incorporating the gradient operator into $\phi_{in}$ such that the model is a function of $\nabla \overline{\mathbf{u}}$. While this preserves invariance with respect to translations of the reference frame, more consideration must be taken for rotation or reflection symmetries. Regarding $\phi_{out}$, the identity would mean that the network outputs each of the tensor components of $\boldsymbol{\tau}_{ml}$. In practice, this can easily lead to numerical instabilities in the simulation during both training and evaluation and is thus not an optimal design choice. We examine several alternatives of $\phi_{in}$ and $\phi_{out}$, which are evaluated in this study.

The first set of models is motivated by tensor basis theory. The use of tensor bases has been extensively leveraged in literature for RANS data-driven turbulence modeling \cite{Ling2016,Milani2021,Berrone2022}. If we assume that the model is an arbitrary function of input tensors, one can construct a finite set of basis tensors whose linear combination via certain weight coefficients is equal to any tensor function of the input, given by:
\begin{equation}
    \boldsymbol{\tau}_{ml} = \sum_n \alpha^{(n)}\mathbf{T}^{(n)},
\end{equation}
where $\mathbf{T}^{(n)}$ are the basis tensors and $\alpha^{(n)}$ are the coefficients.
A common choice of input tensors in turbulence modeling is the strain and rotation rate tensors $\mathbf{S}$ and $\mathbf{R}$, which are the symmetric and anti-symmetric components of the velocity gradient tensor respectively. Pope derived the integrity basis for the $\mathbf{S}$ and $\mathbf{R}$ tensors in two and three dimensions \cite{Pope1975}. The bases in 2D are:
\begin{equation}
    \mathbf{T}^{(0)}=\mathbf{I}, \quad \mathbf{T}^{(1)}=\mathbf{S}, \quad \mathbf{T}^{(2)}=\mathbf{S}\mathbf{R} - \mathbf{R}\mathbf{S}.
\end{equation}

The task of the neural network, then, is to learn the coefficients of the tensor basis functions. It is advantageous to have the coefficients be functions of the invariants of the input tensors, such that the model is ultimately Galilean invariant to rotations and reflections as well as translations. The number of invariants is also finite, $\{\mathbf{S}^2\}$ and $\{\mathbf{R}^2\}$ in 2D, where $\{\mathbf{T}^2\} = T_{ij}T_{ji}$. Ling, et al. used this approach \cite{Ling2016}, also considered a nonlinear eddy viscosity model \cite{Gatski2001}, to model the Reynolds anisotropy tensor using a neural network.

For the model outlined above,
\begin{align}
    \phi_{in}(\overline{\mathbf{u}}) &= \left\{\overline{\mathbf{S}}^2\right\}, \left\{\overline{\mathbf{R}}^2\right\} \\
    f_{\theta}\left(\left\{\overline{\mathbf{S}}^2\right\}, \left\{\overline{\mathbf{R}}^2\right\}\right) &= \alpha\\
    \phi_{out}(\alpha) &= \sum_{n=0}^2 \alpha^{(n)}\mathbf{T}^{(n)},
\end{align}
such that $\boldsymbol{\tau}_{ml} = (\phi_{out}\circ f_\theta \circ \phi_{in})(\overline{\mathbf{u}})$.
Variations of this model are also considered, in terms of the tensor basis functions and thus $\phi_{out}$. The tensor bases above are restricted to symmetric tensors to provide a physically valid stress. Despite this, the basis can be expanded to include asymmetric terms. In 2D, we must only include one additional tensor, $\mathbf{T}^{(3)}=\mathbf{R}$. On the other hand, we can create another model that removes any nonlinear terms in the basis, such that only $\mathbf{T}^{(0)}$ and $\mathbf{T}^{(1)}$ are used. In this case, we recover essentially a linear eddy viscosity model, albeit the effective eddy viscosity is computed from a neural network.

An alternative approach for $\phi_{out}$ can be taken where we disregard the use of tensor bases and instead learn the stress tensor itself, without adapting an existing eddy viscosity model. As mentioned earlier, taking the four components of $\boldsymbol{\tau}_{ml}$ directly from the neural network can lead to numerical issues, and we found that it was impractical to try and train a model in this fashion. We found that it was feasible to split the tensor into its isotropic, deviatoric, and optionally asymmetric components and learn these individually. The stress is then given by
\begin{equation}
    \boldsymbol{\tau}_{ml} = \alpha\mathbf{I}+\mathbf{D}+\mathbf{A},
\end{equation}
where $\mathbf{I}$ is the identity, $\mathbf{D}$ is the deviatoric tensor with two free components, and $\mathbf{A}$ is the anti-symmetric tensor with one free component. The network $f_\theta$ outputs a scalar $\alpha$, the two components comprising $\mathbf{D}$, and possibly one component comprising $\mathbf{A}$. We term this the ``model-free'' (MF) approach.

The secondary aspect to vary is $\phi_{in}$. So far we have considered ANN inputs $\left\{\overline{\mathbf{S}}^2\right\}, \left\{\overline{\mathbf{R}}^2\right\}$. Instead of providing purely the invariants of the tensors $\overline{\mathbf{S}}$ and $\overline{\mathbf{R}}$, we can use the tensors themselves as input, in terms of their two or one free components respectively. While this comes at the expense of rotation invariance, it can provide additional information to the model for more accurate predictions, and the symmetry itself may be learned implicitly by the model.

We refer to various combinations of $\phi_{in}$ and $\phi_{out}$ with several names.
A table of $\phi_{out}$'s and their names are given in Tab. \ref{tab:phi}. A ``-I'' is appended to the name if $\phi_{in}$ provides invariant inputs.

\def\arraystretch{1.3}
\begin{table}[]
\caption{Model names}
\begin{tabular}{r|l}
\textbf{Name} & $\phi_{out}$                                \\ \hline
Linear (LIN)           & $\sum_{n=0}^1 \alpha^{(n)}\mathbf{T}^{(n)}$ \\
Non-linear (NL)            & $\sum_{n=0}^2 \alpha^{(n)}\mathbf{T}^{(n)}$ \\
Non-linear asymmetric (NLA)           & $\sum_{n=0}^3 \alpha^{(n)}\mathbf{T}^{(n)}$ \\
Model-free (MF)            & $\alpha\mathbf{I}+\mathbf{D}+\mathbf{A}$    \\
Model-free symmetric (MFS)           & $\alpha\mathbf{I}+\mathbf{D}$              
\end{tabular}
\label{tab:phi}
\end{table}

\subsection{Deep learning architectures}
The choice of ANN $f_{\theta}$ provides an important inductive bias to the SGS model, particularly with regards to the scales of the flow that are captured by the model. Previous works have focused on using two common architures -- multi-layer perceptrons (MLPs) and convolutional neural networks (CNNs) \cite{Maulik2018,Stoffer2021,Wang2018,Guan2022,List2022}. In this context, MLPs are purely local in nature, in that the evaluation of the SGS stress is parallelized over grid points to produce a model that is only a function of the local strain and rotation rate tensors. The local formulation is supported by by many existing SGS models, including the Smagorinsky model, which is a function of the local strain rate. Despite the success of these local models, it is well known that turbulent dynamics at a point in space are influenced by the surrounding flow as well \cite{ClarkDiLeoni2021}, implying that features of the flow at length scales larger than the grid size may be used to determine the SGS stress. For example, the dynamic Smagorinsky model or other scale-similarity approaches compute model coefficients on-the-fly by filtering the LES flow with a test filter that is larger than the grid filter. The result is that the scales of the flow between the test filter and the LES cutoff are used to estimate the SGS stress. Moreover, it is also common to average model coefficients in space (for instance spanwise direction averaging in channel flow).

The adoption of these non-local, dynamic models in practice motivates the use of CNNs in SGS modeling. The convolutional architecture incorporates information from the surrounding strain and rotation rate quantities through convolutional filters of a specified kernel width. By chaining many of these CNN layers together, the resulting deep model employs a large receptive field that captures a great degree of length scales in the flow. Concretely, the MLP and CNN architectures we use have 6 hidden layers with 64 channels each and ReLU activation functions and one linear output layer. The CNN uses a kernel size of 3 in each layer and includes circular padding to account for periodic boundary conditions. The CNN architecture is the same architecture used by Kochkov, et al. \cite{Kochkov2021}. We note that the input to all networks are size $N_x\times N_y\times M$, where $N_x,N_y$ is the size of the filtered and coarse-grained DNS (FDNS) field or LES grid and $M$ is the number of input channels -- 2 in the case of invariant inputs and 3 in the case of non-invariant inputs. For the MLP, the grid dimensions $N_x,N_y$ are taken as batch axes. 

Thus far, another architectural choice has not yet been explored in the context of 2D turbulence modeling, the Fourier Neural Operator (FNO) \cite{li2021fourier}. FNOs operate in both the real and frequency domain, combining local, pointwise linear operations with pointwise linear transformations of the Fourier modes of the input. From the convolution theorem, the pointwise operations in Fourier space are equivalent to global convolution operations, except the convolutional filter is no longer restricted to be locally supported, which allows for efficient manipulation of the large scales of the input without requiring very deep CNNs.

Assuming an input feature at layer $l$ $h^l \in \mathbb{R}^{N_x\times N_y \times c_{in}}$, where $c_{in}$ is the number of channels, the FNO layer is given by:
\begin{equation}
    h^{l+1}_{nmo} = \sigma(B_{oi}h^l_{nmi} + \mathcal{F}^{-1}(W_{kloi}\mathcal{F}(h^l_{nmi})_{kli})_{nmo}), 
    \label{eqn:fno}
\end{equation}
where $B \in \mathbb{R}^{c_{in} \times c_{out}}$ is a matrix that is contracted along dimension $c_{in}$ of the input in real space and $W \in \mathbb{R}^{k_{max} \times k_{max} \times c_{in} \times c_{out}}$ is contracted along $c_{in}$ pointwise in Fourier space (indexed by $k,l$ for the wavenumbers in each direction) for each wavemode $k$ up to $k_{max}$. The outputs are summed and a pointwise nonlinearity $\sigma$ is applied. The parameters of the layer are the weight tensors $W$ and $B$. We test this approach as well for $f_\theta$, using the hyperparameters presented in the paper \cite{li2021fourier}, which include 4 layers, $k_{max}=12$, and a hidden channel width of 32.

The standard formulation of the FNO is ill-suited in the context of SGS modeling, in that the filters are only applied to the low wavenumbers of the flow. This means that the FNO is capturing the large scales of the flow but disregarding the small scales, which are typically more relevant when determining the SGS stress. We expect, then, that the vanilla FNO will not be able to reproduce the flow spectra accurately. To this end, we design an additional hybrid architecture that combines the benefits of both CNNs and FNOs. The input is first passed through a 2-layer FNO with 4 hidden channels and $k_{max}=10$. The output is projected linearly to a size of 64 hidden channels. Finally, the projection is provided to a 3-layer CNN with otherwise equivalent specifications as the pure CNN architecture to produce the network output quantities. The hyperparameters for this model were chosen through empirical testing and optimized to reduce the memory footprint and computational cost of the architecture. We expect that this model will improve over the other networks because all the length scales in the flow are used to compute the SGS stress.

In summary, we test four different neural network architectures for $f_{\theta}$. The MLP is local and receives no information regarding neighboring stress and rotation rate quantities. The CNN is in some sense semi-local, in that non-local $\overline{\mathbf{S}}$ and $\overline{\mathbf{R}}$ are provided within a specified receptive field. The FNO is global, but only operates on the large scales of $\overline{\mathbf{S}}$ and $\overline{\mathbf{R}}$. Lastly, the hybrid CNN+FNO ultimately captures all the scales of the flow by combining global information from an FNO and non-local information from a CNN.

\subsection{Training}
Model training is accomplished in an end-to-end fashion. Since the solver is implemented in \texttt{JAX}, gradients can be computed with respect to any solver parameter by backpropagating a loss using automatic differentiation. We define the simulation function $\mathcal{S}$ as the numerical algorithm that computes solutions to eq. \ref{eqn:les} along with the incompressibility constraint, using the SGS model given by eqs. \ref{eqn:sgs} and \ref{eqn:sgs_ml}. The solution of the ML-LES model is given by:
\begin{equation}
    \hat{\overline{\mathbf{u}}}^0\dots\hat{\overline{\mathbf{u}}}^t = \mathcal{S}_{ML}(\overline{\mathbf{u}}^0, C_s, \mathcal{M}(\theta)),
\end{equation}
where $\hat{\overline{\mathbf{u}}}^t$ is predicted solution at time $t$, $C_s$ is the Smagorinsky constant, and $\theta$ are the parameters of the ANN inside $\mathcal{M}$.

The target solution is given by the coarse-grained DNS solution field which we hereby denote `subsampled DNS':
\begin{equation}
    \overline{\mathbf{u}}^0\dots\overline{\mathbf{u}}^t = G_S \star \mathcal{S}_{DNS}(\mathbf{u}^0),
\end{equation}
where $\mathcal{S}_{DNS}$ is a high resolution simulation of the unfiltered initial condition and $G_S$ is a coarse-graining operation. Here, we emphasize that the operation $G_S$ \underline{makes no assumption of a low-pass spatial filter} and merely partially observes the high-dimensional state. The objective of the optimization problem is given as:
\begin{equation}
    \min_{C_s,\theta} \mathcal{L}(\overline{\mathbf{u}}^0\dots\overline{\mathbf{u}}^t, \mathcal{S}_{ML}(\overline{\mathbf{u}}^0, C_s, \mathcal{M}(\theta))),
\end{equation}
where we aim to optimize both the Smagorinsky coefficient in $\boldsymbol{\tau}_{smag}$ and the network parameters that produce $\boldsymbol{\tau}_{ml}$ to minimize the loss function $\mathcal{L}$. We take the loss function to be the mean squared error of the trajectories:
\begin{equation}
    \mathcal{L} = \frac{1}{T\cdot N_x \cdot N_y}\sum_{t=1}^{T}\sum_{n=1}^{N_x}\sum_{m=1}^{N_y} (\overline{\mathbf{u}}^t - \hat{\overline{\mathbf{u}}}^t)^2.
\end{equation}

The length of the simulation and temporal sampling of the solution trajectory is of particular interest for training. We would like the loss to remain low for very large $T$, but the memory costs incurred during backpropagation are linear with $T$. Therefore, generally $T_{train}<<T_{eval}$, where $T_{train}$ denotes the length of the simulation during training and $T_{eval}$ denotes the length of simulation during evaluation or testing. As observed by List, et al., $T_{train}$ should be large enough to capture the important temporal scales of the flow, but there are diminshing returns with very large $T_{train}$. We take $T_{train} = 512\Delta t_{DNS} = 64\Delta t_{LES}$, where $\Delta t_{DNS}$ and $\Delta t_{LES}$ are the time steps for DNS and LES simulations respectively. Furthermore, we choose a lower temporal sampling frequency for computing the loss function, such that $\Delta t_{train} = 128\Delta t_{DNS} = 16\Delta t_{LES}$, for a total of 4 snapshots per training trajectory. We found this sufficient for extrapolation to much larger $T_{eval}$, where we tested up to $T_{eval}=131,072\Delta t_{DNS} = 16,384\Delta t_{LES}$, a total of $256\times$ longer than the training simulation.

The models were trained with the Adam optimizer \cite{kingma2017adam} using a decaying learning rate varying from $2\times10^{-3}$ to $4\times10^{-4}$ on 8 V100 GPUs for 300 epochs. In addition, stochastic weight averaging \cite{izmailov2018averaging} was used in the final 20\% of training, where model weights were averaged over the last 20\% of epochs to promote better generalization.

\section{Datasets}
\label{sec:data}

Two-dimensional homogeneous isotropic turbulence (2D-HIT) \cite{Maulik2018} provides an idealized case study to assess model performance. 2D-HIT is a common choice of testbed for training and deploying data-driven models, given its foundations in theory and ability to represent a variety of turbulent flow characteristics. The models are evaluated on several realizations of the same test case, which have been parameterized with respect to Reynolds number and forcing conditions. This yields a sufficiently diverse set of flows on which the generalization ability of the learned SGS models can be demonstrated. For all flow scenarios, the ground truth datasets are obtained from simulating the governing equations in a doubly periodic square domain with $L=2\pi$, discretized on a uniform grid of size $2048\times2048$. The true fields are coarse-grained with $G_S$ to produce target fields with a resolution of $64\times64$.

\def\arraystretch{1.3}
\begin{table}[]
\caption{Dataset parameters}
\label{tab:data}
\begin{tabular}{c|ccccl}
\textbf{Dataset} & $Re$   & $A$ & $k$ & $r$ & $\mathbf{f}$ direction \\ \hline
DE               & 1000   & 0   & 0   & 0   & $\hat{\mathbf{e}}_x$   \\
G1               & 1000   & 1   & 4   & 0.1 & $\hat{\mathbf{e}}_x$   \\
G2               & 30     & 1   & 4   & 0.1 & $\hat{\mathbf{e}}_x$   \\
G3               & $10^5$ & 1   & 4   & 0.1 & $\hat{\mathbf{e}}_x$   \\
G4               & 8000   & 2   & 8   & 0.1 & $\hat{\mathbf{e}}_y$  
\end{tabular}
\end{table}
\begin{figure*}
  \begin{center}
      \includegraphics[trim=2.5cm 2.0cm 2.5cm 2.5cm, clip, width=1.0\textwidth]{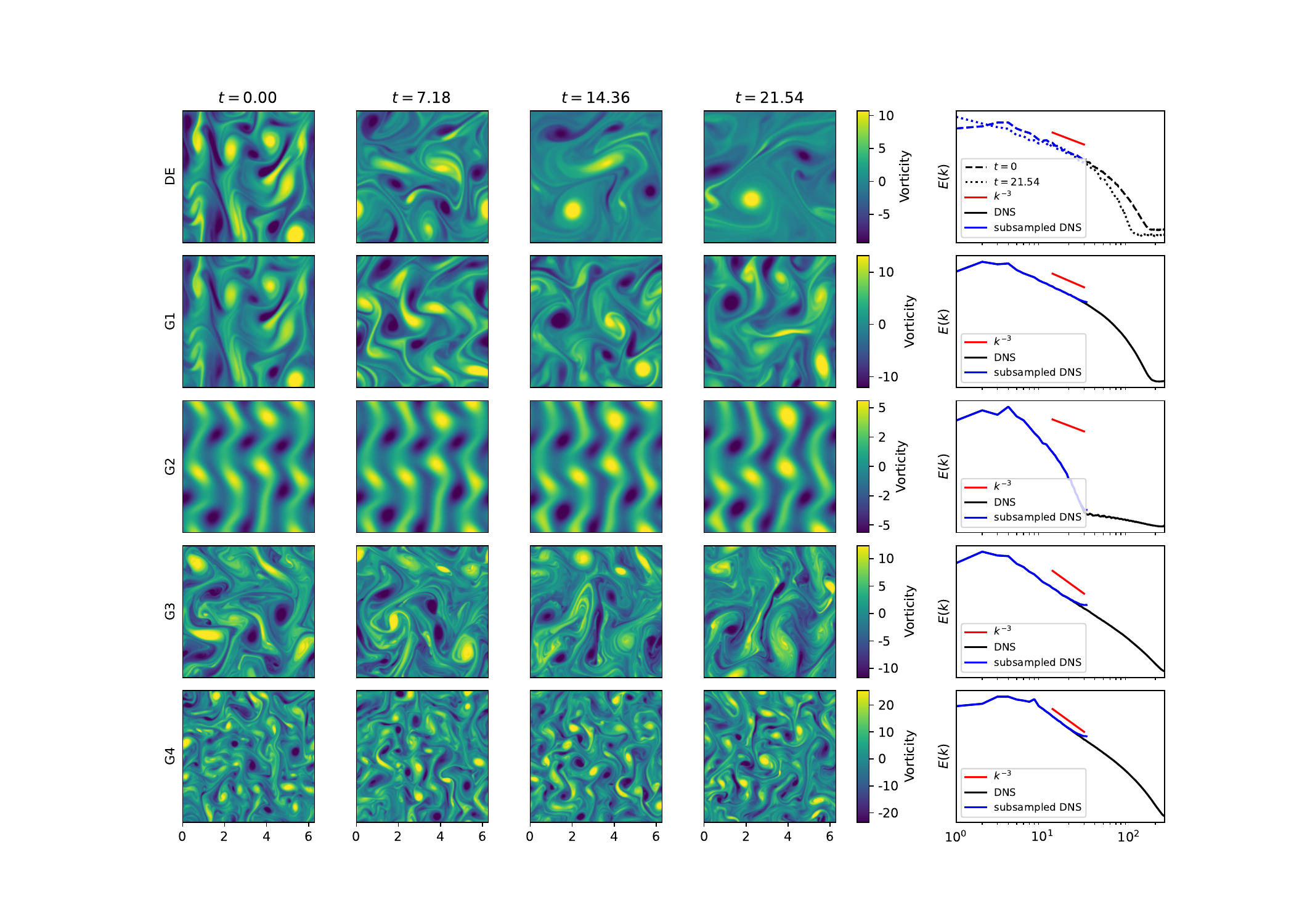}
  \end{center}
  \caption{Example ground truth vorticity field trajectories and energy spectra from each dataset. (DE) Unforced, decaying turbulence at an initial $Re=1000$. This dataset was used for training. (G1) Forced turbulence using a sinusoidal forcing function with wavenumber $k=4$, $Re=1000$. (G2) Low Reynolds number flow, $Re=30$, also forced at $k=4$. (G3) High Reynolds number flow, $Re=10^5$, forced at $k=4$. Given the high $Re$, the ground truth was computed from LES on a $2048\times2048$ grid. (G4) Forced turbulence at a higher wavenumber $k=8$, and a higher $Re=8000$. Additionally, the direction of forcing is rotated $90\deg$ from the previous datasets. All plots show DNS spectra in addition to spectra obtained from subsampled observations of the DNS. The ideal $k^{-3}$ scaling for the inverse cascade in 2D-HIT is also shown for reference.}
  \label{fig:dns}
\end{figure*}

The training dataset, denoted DE, consists of unforced, decaying turbulence at an initial Reynolds number of 1000. A total of 32 trajectories were used for training, each with length $T=16,384\Delta t_{DNS}$. The trajectories were additionally temporally sampled at $\Delta t_{train}$ to produce 128 snapshots per trajectory, further split into data samples of 4 snapshots each for training. For consistency, we fix $\Delta t_{DNS}=2.191\times10^{-4}$ for all datasets, which was sufficient to satisfy the Courant–Friedrichs–Lewy (CFL) criterion with $C_{max}<0.5$. The models are tested on four generalization datasets of forced turbulence. The forcing function $\mathbf{f}$ is given by:
\begin{equation}
    \mathbf{f} = A \mathrm{sin}(ky)\hat{\mathbf{e}}_x - r\mathbf{u},
\end{equation}
and parameterized by the amplitude $A$, wavenumber $k$, and linear drag $r$. We additionally change the forcing direction ($A\mathrm{sin}(kx)\hat{\mathbf{e}}_y$) in one dataset to ensure the models are not overfitting to a particular forcing orientation. A table of the datasets and their relevant parameters are given in Tab. \ref{tab:data}. We show example trajectories from each of the datasets and their energy spectrum in Fig. \ref{fig:dns}, averaged over time for the cases of statistically stationary forced turbulence. The filtered target field's spectrum is also marked. For dataset G3, given the high Reynolds number of $10^5$, DNS is not feasible on a computational grid of $2048\times2048$. Instead, the ground truth data is generated from a high resolution ($2048\times2048$) LES simulation using the Smagorinsky turbulence model, and the target field is coarse-grained from this high-resolution solution.

\section{Results and discussion}
\subsection{Model sweep}
\begin{figure*}
  \begin{center}
      \includegraphics[trim=1.0cm 2.5cm 1.0cm 2.5cm, clip, width=\textwidth]{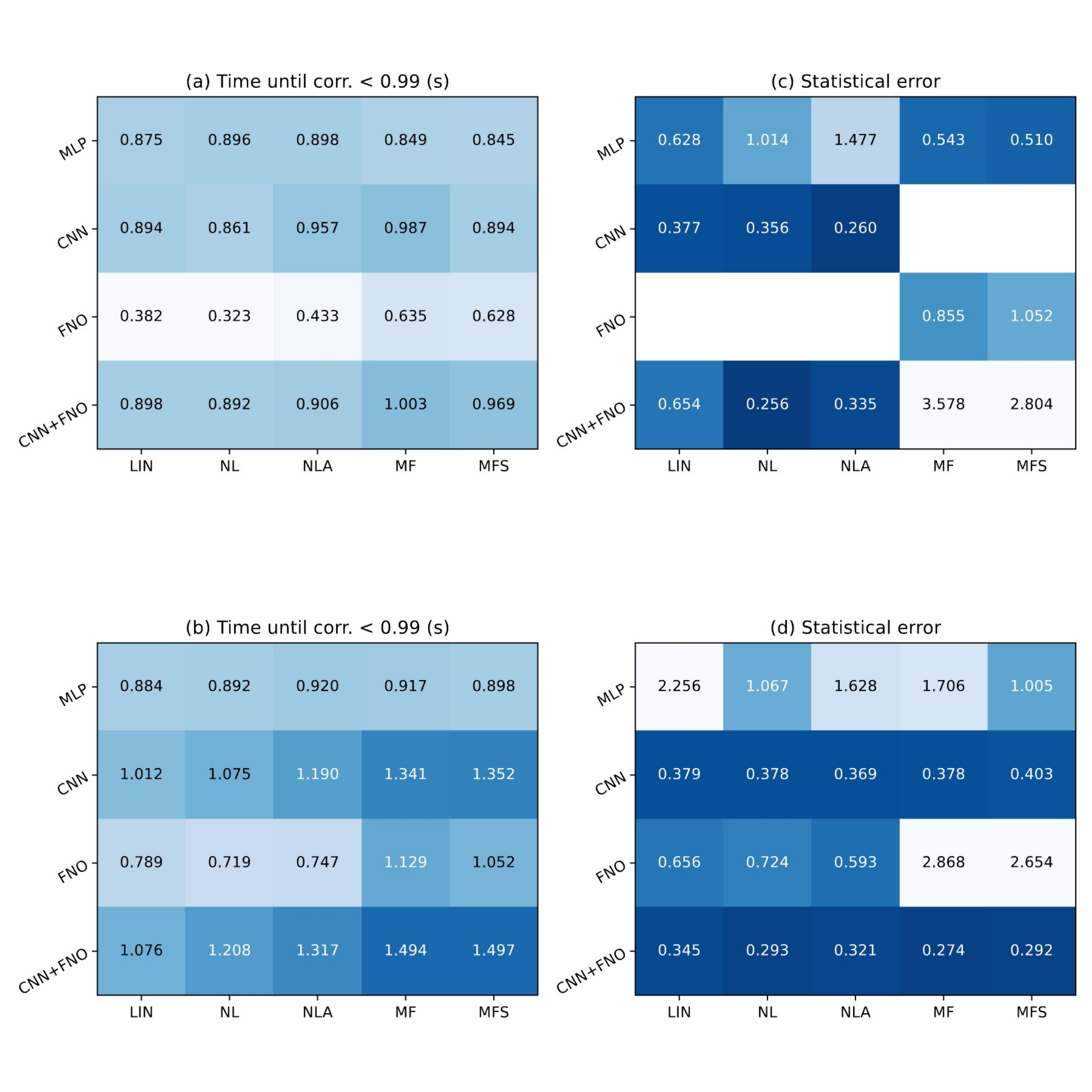}
  \end{center}
  \caption{Ensemble average pointwise and statistical metrics of the predicted solution fields from each of the 40 models tested. (a) and (b) report the pointwise accuracy of invariant input and non-invariant input SGS models respectively, computed as the time before correlation with the ground truth drops below 0.99, where higher values indicate better agreement. (c) and (d) report the statistical error of invariant input and non-invariant input SGS models respectively, computed from the deviation of predicted and true energy spectra, where lower values indicate better agreement. The CNN+FNO-MF and -MFS models achieve the highest pointwise accuracy and the non-invariant input CNN and CNN+FNO models achieve the lowest statistical error. The use of invariant inputs generally results in poorer performance for the non-local architectures, but the reverse is true for the local MLP.}
  \label{fig:all_mods}
\end{figure*}
In the interest of examining all model hyperparameter selections, we train 40 models in a grid search, corresponding to two choices of $\phi_{in}$, four choices of $f_\theta$, and five choices of $\phi_{out}$. We first examine the models' performance on generalization dataset G1 by measuring two quantities to probe the pointwise and statistical accuracy. The pointwise accuracy is determined from the Pearson correlation coefficient of the predicted velocity field with respect to the ground truth FDNS velocity field. We report the time at which the correlation coefficient drops below 0.99, where higher values indicate greater pointwise accuracy. Pointwise accuracy is useful is certain situations, e.g. short-term weather forecasting, where it is valuable to match the true flow field exactly despite the chaotic nature of the system. In other scenarios, if pointwise accuracy cannot be guaranteed for long-term predictions, it is sufficient to ensure that the statistical behavior of the coarse-grained system is consistent with the ground truth statistics. Here, we report a statistical error based on the time-averaged energy spectrum $E(k)$ of the flow, which displays a characteristic scaling behavior in the inertial range of the spectrum.
The error is computed as:
\begin{equation}
    \left(\frac{1}{K}\sum_{k=1}^K \mathrm{log}\left(\frac{\hat{E}(k)}{E(k)}\right)^2\right)^{1/2},
\end{equation}
where $\hat{E}(k)$ is the predicted energy spectrum and $E(k)$ is the energy spectrum of the DNS field. Importantly, the errors at all wavenumbers are weighted equally in this formulation, as opposed to e.g. mean-squared error (MSE), which would favor accuracy at the higher energy-containing low wavenumbers. We also perform the time-averaging after the prediction has decorrelated with the ground truth trajectory.

Fig. \ref{fig:all_mods} shows the pointwise and statistical measures on the G1 dataset of all models tested. The left two grids and right two grids display the pointwise accuracy and statistical error respectively. Examining the pointwise accuracy, we find that the use of invariant inputs (the upper row of models) negatively impacts accuracy, although the impact is nearly insignificant for the MLP models. Overall, the CNN and CNN+FNO architectures achieve the highest accuracy, particularly in combination with MF or MFS. Looking at the statistical error, again the CNN and CNN+FNO architectures achieve the lowest error, but only in conjunction with non-invariant inputs. Many of the non-local architectures with invariant inputs resulted in unstable simulations that diverged before time-averaging could be performed; as a result, the error for these models is not reported. In contrast, the use of invariant inputs is in fact beneficial for the purely local MLP models. We conclude that the orientation of the velocity gradient tensor, provided by the non-invariant inputs, is useful and perhaps necessary information to provide when constructing non-local data-driven SGS models, but harmful when given to local SGS models. We also see that the FNO-only architecture generally performs most poorly, likely because it cannot capture the small-scale behavior of the flow that is relevant for determining the SGS stress. Ultimately, the CNN+FNO-MFS model achieves the best combination of pointwise and statistical accuracy, and we select this as the best performing model for subsequent analysis.

\begin{table}[]
\caption{Evaluation time for fixed forecast length}
\begin{tabular}{r|l}
Model   & Time (s) \\ \hline
DNS     & $137\pm0.490$       \\
SMAG    & $2.28\pm0.071$      \\
MLP     & $3.35\pm0.129$      \\
CNN     & $5.32\pm0.148$      \\
FNO     & $6.81\pm0.093$      \\
CNN+FNO & $5.75\pm0.016$     
\end{tabular}
\label{tab:time}
\end{table}

We report the evaluation times of several models for a fixed forecast length of $2^{17}\Delta t_{DNS}$ in Tab. \ref{tab:time}. Since the ANN architecture was determined to be the most significant contributor to variable runtime, we compare the four architectures against the ground truth DNS and the baseline Smagorinsky closure using a coefficient of 0.172. The MLP is cheapest to compute, while the FNO is the most expensive. Both the CNN and CNN+FNO had similar runtimes, a little over double the baseline model. 

\subsection{Spectral analysis of architectures}

To further understand the impact, mechanisms, and spectral behavior of the neural network architectures, we can plot the radially averaged power spectrum of the network outputs. The power spectrum is computed from integrating the squared 2D Fourier transform of the network outputs over constant magnitude wavenumbers $k=\sqrt{k_x^2+k_y^2}$. The spectrum can assist in revealing which length scales in the flow each of the network architectures is responding to. We start by isolating $f_\theta$ from the rest of the components of the SGS model. To provide a consistent input baseline, we probe the spectral response to white noise, which has a uniform power spectrum. We examine MFS trained models with different architectures and plot $\alpha$, the isotropic component of the SGS stress, and $|\mathbf{D}|$, the magnitude of the deviatoric component of the SGS stress. While these outputs have no physical significance regarding the true SGS stress considering the synthetic input, they can provide insight into the scales captured by the model. We would expect, for example, an MLP to produce entirely flat power spectra as it only operates on local quantities and contains no interactions between length scales. We also expect the FNO power to be uniform at high wavenumbers because the FNO cannot capture any length scales smaller than $L/k_{max}$, where $k_{max}=12$.

\begin{figure}
  \begin{center}
      \includegraphics[trim=0 1.05cm 0 0,clip,width=\linewidth]{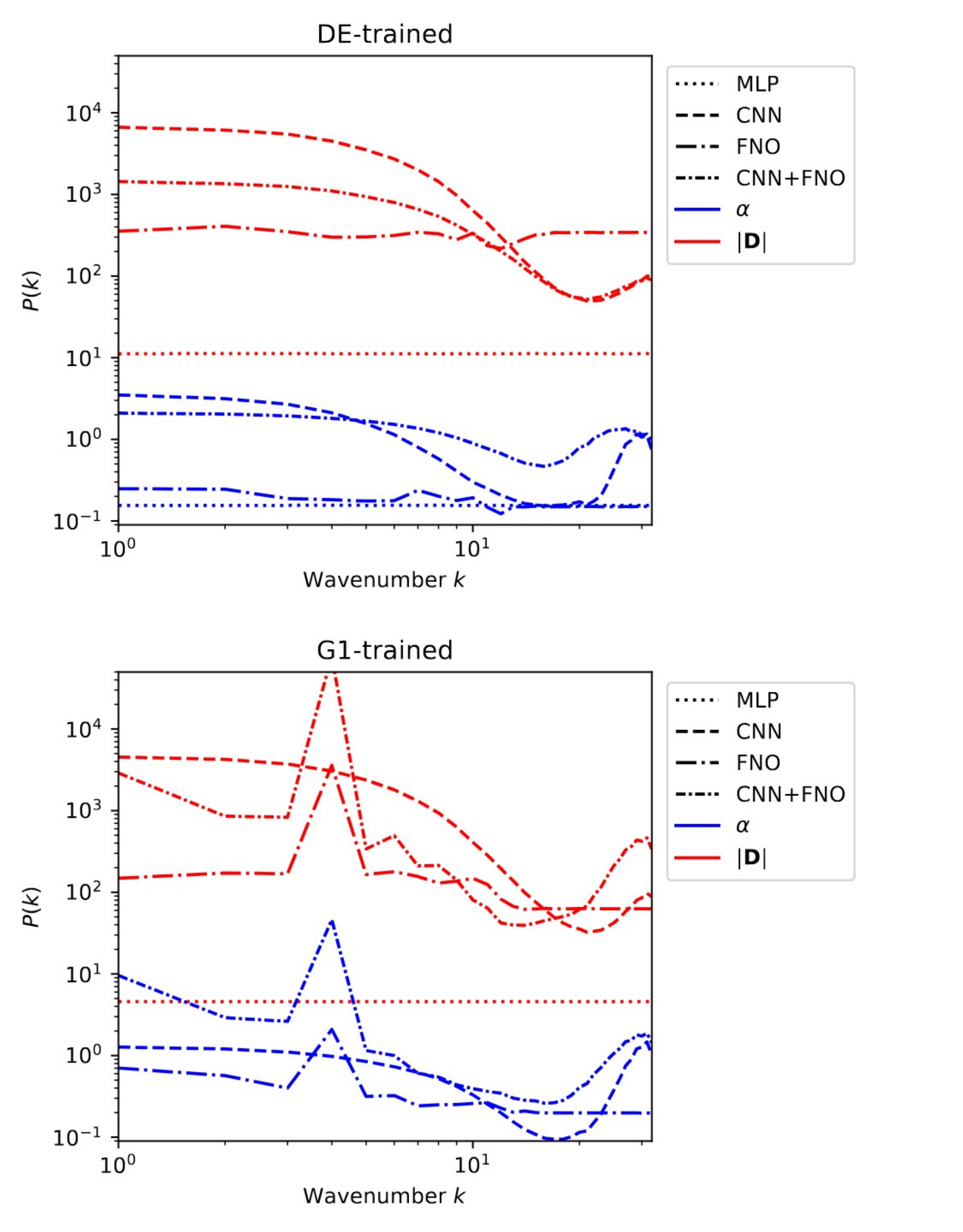}
  \end{center}
  \caption{Examination of the power spectral response of neural architectures to white noise. Outputs of MFS model networks trained on (a) the DE dataset and (b) the forced G1 dataset are shown, with the isotropic component $\alpha$ and magnitude of the deviatoric component $|\mathbf{D}$| visualized. In (a) while the MLP outputs are flat at all wavenumbers and the FNO outputs are flat at high wavenumbers, the CNN and CNN+FNO models display more significant variation at the high wavenumbers, leveling out at the low wavenumbers. In (b) the FNO and CNN+FNO display sharp peaks at the forcing wavenumber, indicating their ability to capture (or overfit to) large-scale features.}
  \label{fig:noise}
\end{figure}
\begin{figure*}
  \begin{center}
      \includegraphics[width=\linewidth]{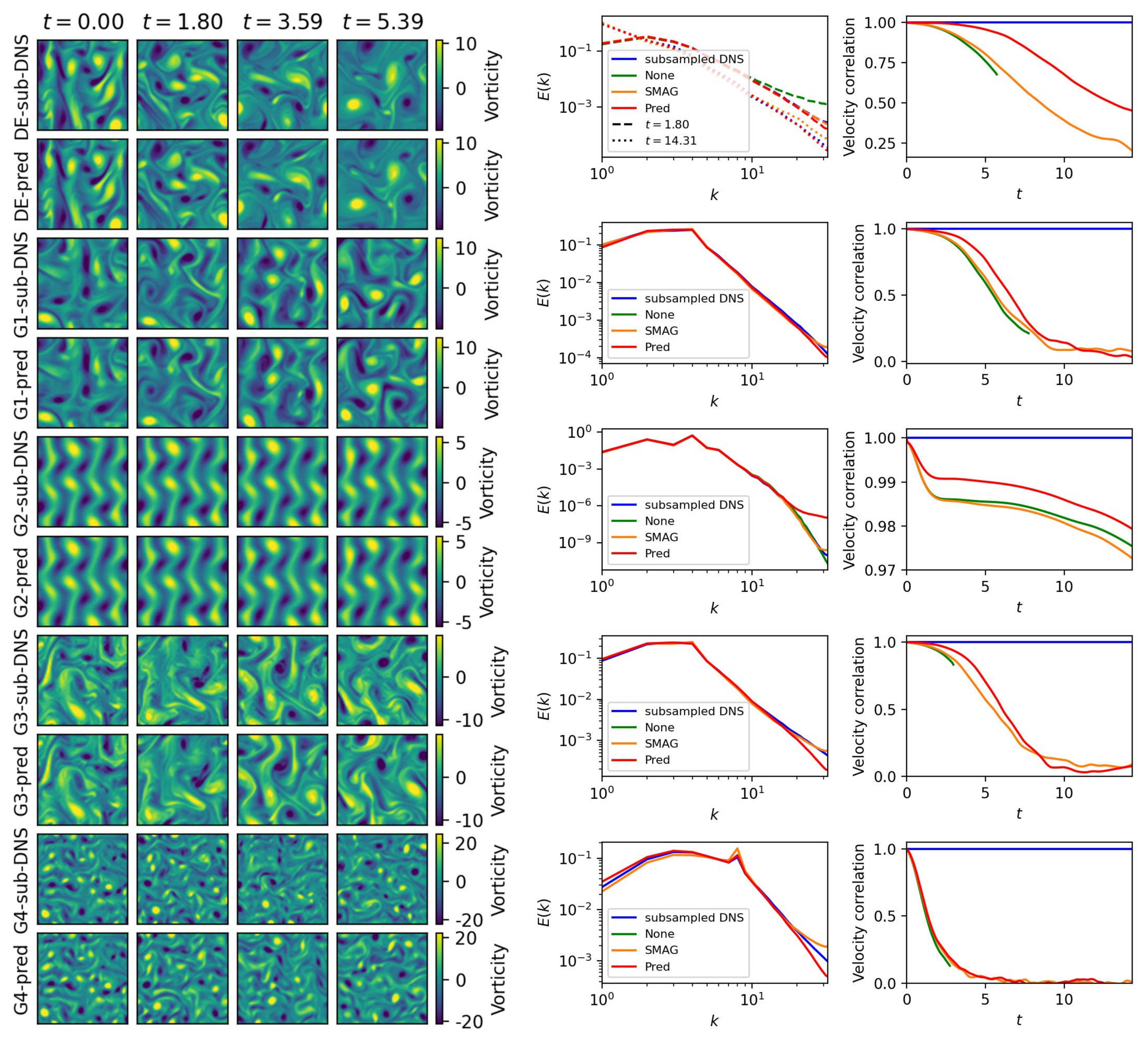}
  \end{center}
  \caption{Summary of predictions from the CNN+FNO-MFS model. The left column displays an example predicted vorticity trajectory for each dataset compared with the true field. The right column shows the ensemble average energy spectra and velocity correlation over time for each dataset. Comparisons to baseline SMAG and no turbulence model are included. The CNN+FNO-MFS model improves over the baseline in both metrics for nearly every case. Only in the low Reynolds number dataset G2 does the model overpredict energy at the highest wavenumbers.}
  \label{fig:generalization}
\end{figure*}

Fig. \ref{fig:noise}(a) shows the power spectra of various networks that have been trained on the DE dataset. We see that our expectations regarding the MLP and FNO are validated, in that the MLP power is uniform for all wavenumbers and the FNO power is uniform at wavenumbers greater than 12. On the other hand, we see that both the CNN and CNN+FNO networks display significantly more variation in the power spectra at high wavenumbers, but are much more uniform at low wavenumbers. This indicates that the CNN models are mostly responding to the smallest length scales in the flow, which is consistent with the network design. Considering the accuracy of the CNN and CNN+FNO models compared to the FNO models, it is clear that the smallest length scales in the flow are most important for determining the SGS stress, again aligned with our expectations regarding turbulence. 

We can contrast this with the spectral response of networks trained on the G1 dataset, which is forced at a wavenumber of 4. The FNO-containing models display a very clear peak at this forcing wavenumber, which indicates that they are overfitting to the training set. Because the FNO operates directly in spectral space, it can easily capture large scale behavior, such as external forcing. The CNN, however, has no such peak since its small receptive field limits its ability to respond to the low frequency modes of the flow. There is a tradeoff, then, in terms of incorporating very non-local, i.e. large scale, features in the SGS model; while this will lead to improved accuracy on in-distribution test cases, it will likely hamper performance on out-of-distribution generalization cases.

\subsection{Generalization to new flows}

A turbulence model's value is strongly tied to its versatility. A significant reason the Smagorinsky model is still widely used in practice today despite its simplicity is that it is effective in a very wide variety of flow configurations. Oftentimes data-driven models cannot match the flexibility of theory-based models because they are trained only on a specific distribution of data. Here, we demonstrate the versatility of our model design by applying, without retraining, the same DE-trained SGS model to each of the four generalization datasets. Fig. \ref{fig:generalization} shows a summary of the model's performance on all datasets, where we have selected the CNN+FNO-MFS model as the data-driven model to test. An example trajectory for each dataset, both the true field and the predicted field, is shown on the left. We observe qualitatively accurate predictions for over 3000 LES time steps in nearly all cases, except for G4, where the eddy turnover time is much shorter due to the smaller characteristic length scale. On the right, we show the energy spectra and transient velocity field correlation, comparing statistics from the true filtered field, no model, Smagorinsky model, and our ML model. The ML model matches the true energy spectrum and improves over the Smagorinsky baseline in nearly all datasets. The exception is the very low Reynolds number case, G2, where the model overestimates the energy at the smallest length scales. This is characteristic of the use of turbulence models when DNS is capable of resolving the flow, exemplified by a similar trend in the Smagorinsky model compared to no model. We note that the energy at these high wavenumbers is very low, and there is no significant impact on the pointwise accuracy. The velocity correlation plots show that the ML model achieves greater pointwise accuracy than the Smagorinsky model for every dataset, with the greatest improvement predictably seen on the DE dataset.

\subsection{Comparison with \textit{a-priori} approach}

\begin{figure}
  \begin{center}
      \includegraphics[trim=0 0.5cm 0 0,clip,width=\linewidth]{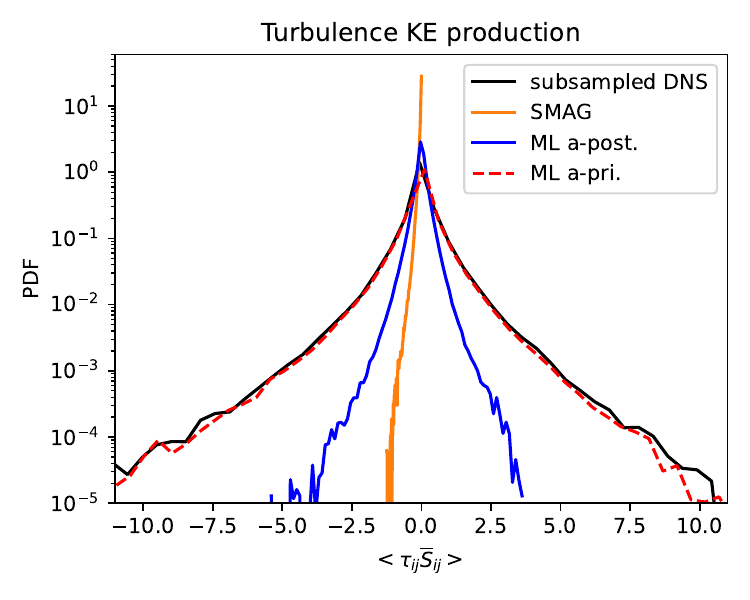}
  \end{center}
  \caption{The \textit{a-priori} computed turbulence kinetic energy production PDF is visualized for the \textit{a-priori} trained and \textit{a-posteriori} trained ML models, as well as the true DNS and SMAG baseline. While SMAG is purely dissipative, the ML models can replicate backscatter, indicated by both positive and negative values. }
  \label{fig:kep}
\end{figure}
\begin{figure}
  \begin{center}
      \includegraphics[width=0.85\linewidth]{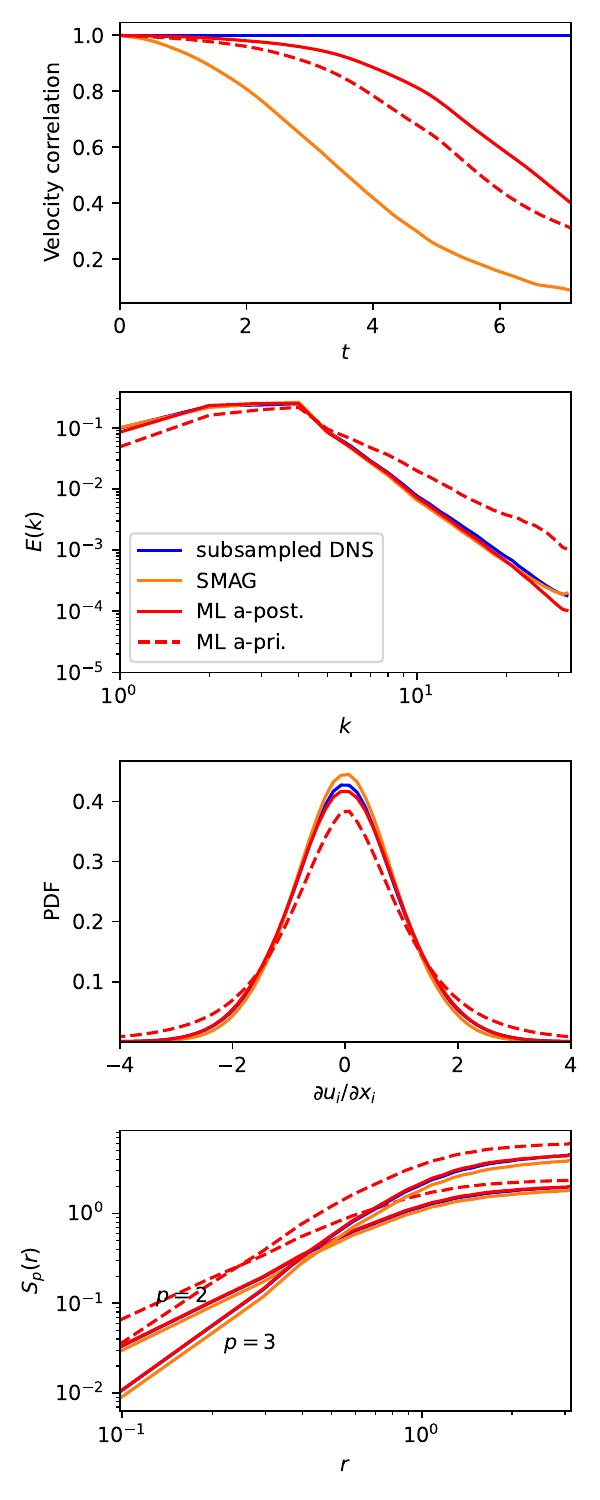}
  \end{center}
  \caption{Selected \textit{a-posteriori} statistics from results on G1 dataset comparing \textit{a-priori} trained and \textit{a-posteriori} trained ML models. (a) velocity correlation with FDNS. The \textit{a-priori} model displays only marginal improvement over SMAG. (b) averaged energy spectra of the flow. While there is near perfect agreement with the DNS from the \textit{a-posteriori} trained model, the \textit{a-priori} model displays more deviation from the ground truth. (c) PDF of velocity gradients. The \textit{a-priori} model has a slightly narrower distribution than the DNS. (d) second and third order structure functions. Both SMAG and the \textit{a-posteriori} model display no noticeable disagreement with the DNS, in contrast to the \textit{a-priori} model.}
  \label{fig:apost}
\end{figure}

Lastly, we compare our end-to-end, differentiable physics approach to learning SGS models with a more conventional offline learning approach. In the \textit{a-priori} setting, the learning task is to match the SGS stress field computed from the true DNS solution using an arbitrary filter, which can be accomplished without integrating the ML model into a simulation during training. This approach eliminates the initial overhead of developing a differentiable solver and the training cost of backpropagating through the simulation, but there are implications in terms of \textit{a-posteriori} accuracy. When deployed online, numerical stability is typically lost and the choice of numerical scheme may lead to unexpected \textit{a-posteriori} performance. Usually, this instability is closely associated with the choice of a filtering operation $G_\Delta^\ast$ that approximates some true, but unknown, $G_\Delta$.

To compare our differentiable physics approach with an \emph{a priori} trained model, we must first make an assumption for a low-pass spatial filter to compute the target subgrid stress. This procedure is explained in the following. We train an equivalent CNN+FNO-MFS model in an \textit{a-priori} setting and evaluate both \textit{a-priori} and \textit{a-posteriori} accuracy on the G1 dataset. The SGS model definition is unchanged, given by eqs. \ref{eqn:sgs} and \ref{eqn:sgs_ml}, but the loss function is more straightforward, simply:
\begin{equation}
    \mathcal{L} = MSE(\hat{\boldsymbol{\tau}}, \boldsymbol{\tau}_{FDNS}),
\end{equation}
where $\hat{\boldsymbol{\tau}}$ is the predicted SGS stress and $\boldsymbol{\tau}_{FDNS}$ is the SGS stress computed from the DNS field using eqn. \ref{eqn:t_dns} with an approximate filter $G_\Delta^\ast$. Here we utilize a popular filtering technique\cite{Kochkov2021} to compute the mean of the values that lie on the face of the new control volume in a given direction and discard the values that do not lie in that  particular direction. This ensures that zero divergence properties are propagated to the downsampled fields too. In such a manner, we can filter ground truth DNS fields which can be followed by subsampling to the coarse grid. Ultimately, we can generate target subgrid stress data on the coarse-grid. Note that different choices of filters may lead to different $\boldsymbol{\tau}_{FDNS}$ computations.

Fig. \ref{fig:kep} shows the probability distribution function (PDF) of the \textit{a-priori} computed turbulent kinetic energy production term, given by $<\tau_{ij}\overline{S}_{ij}>$, computed from the DNS data with the aforementioned approximate filter $G_\Delta^\ast$, the Smagorinsky model, the \textit{a-posteriori} trained ML model, and the \textit{a-priori} trained ML model. First, we notice that because the Smagorinsky model is purely dissipative, the production term is always negative. In addition, the distribution is much narrower than the true distribution. On the other hand, the \emph{a priori} trained predictions match this approximately generated PDF exactly. The \textit{a-posteriori} model's PDF is generated by simply computing the turbulent kinetic energy production term using the grid-resolved quantities during LES deployment. Surprisingly, this PDF is dramatically different from that computed using the approximately filtered data in \textit{a priori}. \textbf{This plot demonstrates that a-priori trained machine learning closures effectively solve \emph{the wrong problem} by making extremely strong assumptions about the nature of the true filter which is unknown}. Moreover, it also becomes clear that the differentiable physics paradigm can be interpreted as an inverse problem solve, that discovers the nature of the true subgrid stress in the absence of assumptions about the filter.

To further validate this claim - we analyze other statistical measures in Fig. \ref{fig:apost}. The \emph{a-priori} trained model as-is was found to be unstable and as such, it was necessary to introduce a clipping method, similar to what was suggested by Maulik, et al. \cite{Maulik2018}, to ensure the SGS model was purely dissipative. Fig. \ref{fig:apost}(a) shows a comparison of the average energy spectra, Fig. \ref{fig:apost}(b) shows the PDFs of velocity gradients, Fig. \ref{fig:apost}(c) shows the second- and third-order structure functions, and Fig. \ref{fig:apost}(d) shows the velocity field correlation over time. For all measures, we see deviations in the \textit{a-priori} model from the filtered DNS field, often greater than those from the Smagorinsky model, while the \textit{a-posteriori} model shows near perfect agreement. We find that although there are some marginal benefits to training SGS models in an \textit{a-priori} setting, these are substantially outweighed by the improvements offered from end-to-end differentiable solvers.

\section{Conclusions}
In this work, we seek to learn subgrid-scale models for LES using differentiable turbulence, by leveraging a differentiable CFD solver paired with deep learning to minimize an \textit{a-posteriori} loss function. In LES, the large scales of the flow are directly resolved while the small scales are modeled; the effect of the unresolved small scales on the large scales must be accounted for through the use of an SGS model to produce accurate solution trajectories. Filtering out the small scales of the solution results in an additional effective SGS stress within the governing equations, which we approximate with a deep learning model. Given that the goal of LES is to realize simulations that can accurately predict the filtered and coarse-grained true solution field, we use an adjoint solver implementation to optimize the \textit{a-posteriori} solution directly, instead of the optimizing the SGS stress computed from a resolved simulation \textit{a-priori}.

Models are evaluated on a variety of two-dimensional turbulent flows. While models are trained on decaying turbulent flows with an initial Reynolds number of 1000, they are tested on turbulent flow with different forcing conditions and Reynolds numbers that span four orders of magnitude. We find that our approach can generalize well to each test case, and improves over the Smagorinsky closure baseline at a fixed coefficient of 0.172 in every scenario. Model design is investigated in detail, including the choice of ANN inputs, outputs, and architecture. We observe that non-linear eddy viscosity model approaches are not as effective as directly learning the mapping between filtered velocity gradients and SGS stresses, which can be achieved through decomposition of the tensors into isotropic, deviatoric, and anti-symmetric components. Furthermore, the non-locality of the network architecture strongly influences model performance, where it is most critical to include small scale interactions near the grid size. The insertion of large scale behavior through an FNO boosts the accuracy of predictions, but may come at the expense of generalization ability depending on the training data.

We demonstrate the benefits of a fully differentiable solver by outlining the improvements in results from \textit{a-posteriori} learning when compared to \textit{a-priori} learning. Given the accuracy that can be achieved even at very large coarse-graining factors, we believe hybrid deep learning-solver algorithms are a very encouraging avenue for cheap and accurate CFD simulations. The incorporation of physics through the solver allows significantly more potential for out-of-distribution performance relative to pure ML algorithms. While this preliminary study has been restricted to idealized flows with a specific discretization, we anticipate future work to investigate the application of this approach to more complex wall-bounded flows on unstructured grids.

\section{Data availability}

Data and software utilized for this study will be made available by the corresponding author on reasonable request. 

\section{Acknowledgements}
This material is based upon work supported by the National Science Foundation Graduate Research Fellowship under Grant No. DGE 1745016 awarded to VS.  The authors from CMU and Michigan acknowledge the support from the Technologies for Safe and Efficient Transportation University Transportation Center, and Mobility21, A United States Department of Transportation National University Transportation Center. This work was supported in part by Oracle Cloud credits and related resources provided by the Oracle for Research program.
This work was supported by the U.S. Department of Energy (DOE), Office of Science, Office of Advanced Scientific Computing Research (ASCR), under Contract No. DEAC02–06CH11357, at Argonne National Laboratory and Penn State University. We also acknowledge funding support from ASCR for DOE-FOA-2493 ``Data-intensive scientific machine learning''

\bibliography{bib}

\end{document}